\documentclass[aps,pra,twocolumn,footinbib,superscriptaddress]{revtex4-1}
\usepackage[T2A]{fontenc}
\usepackage[utf8]{inputenc}
\usepackage{amsmath}
\usepackage{amssymb}
\usepackage{color,xcolor}
\usepackage{natbib}
\usepackage[colorlinks=true,bookmarks=false,citecolor=blue,urlcolor=blue]{hyperref}
\usepackage{bm}
\usepackage{epsfig}
\usepackage{verbatim} 
\usepackage{morefloats}
\usepackage{graphicx}
\usepackage{bibentry}
\usepackage[mathscr]{euscript}
\usepackage{enumerate}
\usepackage{ bbold } 
\usepackage{ upgreek }
\usepackage{subfigure}
\usepackage{amsmath}
\usepackage{amssymb}

\usepackage{bm}

\def\dfrac{\displaystyle\frac}  
\newcommand{\eps}{\varepsilon}

 \usepackage{lipsum}
 
\renewcommand{\phi}{\varphi}

\begin{document}

\author{\firstname{Ekaterina O.} \surname{Smolina}}
\affiliation{Department of Control Theory, Nizhny Novgorod State University, Gagarin Av. 23, Nizhny Novgorod, 603950 Russia}

\author{\firstname{Lev A.} \surname{Smirnov}}
\affiliation{Department of Control Theory, Nizhny Novgorod State University, Gagarin Av. 23, Nizhny Novgorod, 603950 Russia}

\author{\firstname{Daniel} \surname{Leykam}}
\affiliation{Centre for Quantum Technologies, National University of Singapore, 3 Science Drive 2, Singapore 117543}

\author{\firstname{Franco } \surname{Nori}}
\affiliation{Theoretical Quantum Physics Laboratory, Cluster for Pioneering Research, RIKEN, Wakoshi, Saitama 351-0198, Japan}
\affiliation{Center for Quantum Computing (RQC), RIKEN, Wako-shi, Saitama 351-0198, Japan}
\affiliation{Physics Department, University of Michigan, Ann Arbor, MI 48109-1040, USA}

\author{\firstname{Daria A.} \surname{Smirnova}}
\affiliation{Theoretical Quantum Physics Laboratory, Cluster for Pioneering Research, RIKEN, Wakoshi, Saitama 351-0198, Japan}
\affiliation{Research School of Physics, Australian National University, Canberra, ACT 2601, Australia}

\title{Identifying topology of leaky photonic lattices with machine learning}

\begin{abstract}
We show how machine learning techniques can be applied for the classification of topological phases in leaky photonic lattices using limited measurement data. We propose an approach based solely on bulk intensity measurements, thus exempt from the need for complicated phase retrieval procedures. In particular, we design a fully connected neural network that accurately determines topological properties from the output intensity distribution in dimerized waveguide arrays with leaky channels, after propagation of a spatially localized initial excitation at a finite distance, in a setting that closely emulates realistic experimental conditions.
\end{abstract}

\maketitle

\section{Introduction}

Machine learning holds great promise for solving a variety of problems in nanophotonics. Rather than attempting to model the system of interest exactly from first principles (e.g., by solving Maxwell's equations), machine learning techniques aim to discover or reproduce key features of a system by optimizing parametrized models using a set of training data~\cite{RevModPhys.91.045002}. A trained model can often predict the properties of a device faster than conventional simulation techniques~\cite{Wiecha2020,Chen2022}. Machine learning can also be used to solve the inverse problems of how to design a nanophotonic structure with desired functionalities, and how to reconstruct the parameters of a device using indirect measurements~\cite{Melati2019,malkiel_mrejen_wolf_suchowski_2020,SoBadloeNohBravoAbadRho+2020+1041+1057,Wiecha:21,LiuLiuZhouLiYouQiuWu+2023+1943+1955}. The latter is particularly important for nanophotonic devices, since structural parameters may differ substantially from the nominal design due to fabrication imperfections.

Recently developed topological photonic systems 
provide a useful testbed for better understanding the capabilities and limitations of machine learning approaches in nanophotonics~\cite{doi:10.1080/23746149.2022.2046156,Pilozzi2018}. Topological photonic structures host robust edge states which are protected against certain classes of fabrication imperfections. This robustness is explained by the bulk-boundary correspondence, which relates the existence of localized boundary modes to nonlocal topological invariants expressed as integrals of a connection or curvature of the bulk modes~\cite{ozawa2019topological}. While the direct measurement of a topological invariant entails the reconstruction of both the intensity and phase profiles of the bulk modes of a structure, machine learning models can perform supervised classification of topological phases using a limited set of observables~\cite{doi:10.1080/23746149.2022.2046156}.

In general, the performance of machine learning depends on both the quality and quantity of the data used to train the model. Supervised learning approaches, such as deep neural networks, typically require a huge quantity of labelled training data, which may be hard to come by. This has motivated recent interest in the use of unsupervised learning techniques such as manifold learning, which do not require labelled training data to distinguish topological phases~\cite{Rodriguez-Nieva2019,PhysRevLett.124.226401,PhysRevLett.125.127401,PhysRevB.102.134213,PhysRevLett.130.036601}. Broadly speaking, these techniques are sensitive to sharp changes to observables that occur in the vicinity of topological phase transition points, and thus perform best when one has access to measurements from a large set of different model parameters, which is most feasible when the parameter controlling the phase transition is continuously tunable~\cite{PhysRevLett.125.127401}.

The above methods also rely on prior knowledge of the characteristics of the physical system (such as its sizes, its internal structure and the parameters of the initial excitation), therefore, being not in line with a realistic experimental framework. Data quality and feature selection can have a significant impact on the machine learning-based reconstruction of topological phase diagrams~\cite{PhysRevResearch.2.023283}. For example, missing data arising from incomplete measurements or local perturbations to the data can act as adversarial attacks that fool neural network-based classifiers of topological phases into making incorrect predictions~\cite{Zhang2022}. The existence of adversarial examples highlights the importance of taking platform-specific uncertainties and disorder into account in the selection and design of machine learning classifiers of topological phases. 

The aim of this study is to investigate how common obstacles encountered in the characterization of nanophotonic devices -- disorder, imperfect alignment, and access to a limited set of output observables -- affect the performance of machine learning-based classification and clustering methods for topological phases. Specifically, we focus on the case of one-dimensional waveguide arrays which have provided a versatile platform for the investigation of topological effects in nanophotonics~\cite{8848487,https://doi.org/10.1002/adpr.202100010,PhysRevResearch.4.033222}, considering the problem of predicting the existence or absence of edge states based on bulk intensity measurements. First, we show that while curated input data can improve the performance of clustering, ambiguity in the training data (in the form of uncertainty in the alignment of the input waveguide) leads to incorrect cluster assignments, requiring the use of supervised learning techniques. We compare the performance of several supervised classification models, including a convolutional neural network, demonstrating the ability to predict the existence of different edge state configurations with high accuracy using bulk intensity measurements. Finally, we show the feasibility of transfer learning for sufficiently weak disorder strengths, i.e. maintaining accurate predictions of topological edge states using a model trained on disorder-free data. Our numerical results reveal the feasibility using machine learning techniques to distinguish nanophotonic topological phases using incomplete measurements.

The outline of this article is as follows: Section~\ref{sec:model} reviews the properties of the leaky Su-Schrieffer-Heeger (SSH) tight binding model and introduces the datasets which will be used in our study. Section~\ref{sec:tsne} presents the results of unsupervised clustering according to the edge state configuration using the t-distributed stochastic neighbor embedding (t-SNE) method. We compare the performance of different supervised learning techniques in Sec.~\ref{sec:supervised}. As an example of the feasibility of transfer learning we consider in Sec.~\ref{sec:disorder} the classification performance for disordered waveguide arrays. We conclude with Sec.~\ref{sec:conclusion}. The Supplementary Materials contain additional details on the tight binding model parameters, training data, and the employed machine learning models.

\section{Model and dataset preparation}
\label{sec:model}

We consider light propagation in waveguide arrays governed by the paraxial wave equation,
\begin{equation}
i \dfrac {\partial \mathcal{E}}{\partial z} + \dfrac{1}{2 k_0} \Delta_{\bot} \mathcal{E} + \dfrac{ k_0 n_L(\boldsymbol{r}_{\perp}) }{n_0} \mathcal {E} = 0\:,
\label{eq:paraxial}
\end{equation}
where $\mathcal{E}$ is the envelope of the optical wavepacket propagating along the $z$ (waveguide) axis, $\boldsymbol {r}_{\bot} = (x,y)$ are the transverse coordinates, $k_0 = {2 \pi n_0}/{\lambda}$ is the wave number, $n_L (\boldsymbol {r}_{\bot}) $ is a perturbation of the refractive index forming the waveguide lattice, and $n_0$ is the background refractive index of the medium.

Formally, the final state after a propagation distance $L$ can be obtained by projecting the input ($z=0$) state $\mathcal{E}(0,\boldsymbol{r}_{\perp})$ onto the propagation-invariant modes of the array $\phi_n(\boldsymbol{r}_{\perp})$ with propagation constant $\beta_n$, i.e.
\begin{equation}
\mathcal{E}(L,\boldsymbol{r}_{\perp}) = \sum_n A_n e^{-i \beta_n L} \phi_n(\boldsymbol{r}_{\perp}),
\end{equation}
where $A_n = \int d\boldsymbol{r}_{\perp} \phi_n^*(\boldsymbol{r}) \mathcal{E}(0,\boldsymbol{r}_{\perp})$ are the amplitudes of the modes excited at the input ($z=0$). The intensity of the final state
\begin{equation}
    |\mathcal{E}(L,\boldsymbol{r}_{\perp})|^2 = \sum_{mn} A_n A_m^* \phi_n(\boldsymbol{r}_{\perp})\phi_m^*(\boldsymbol{r}_{\perp}) e^{i (\beta_m - \beta_n)L}
\end{equation}
is sensitive to both the modal excitation amplitudes $A_n$ and the propagation length $L$, so intensity measurements at a single $L$ are generally insufficient to uniquely reconstruct the modal profiles, propagation constants, and topological invariants of the system.

\begin{figure}[ht!]
	\centering
\includegraphics[width=8.4cm]{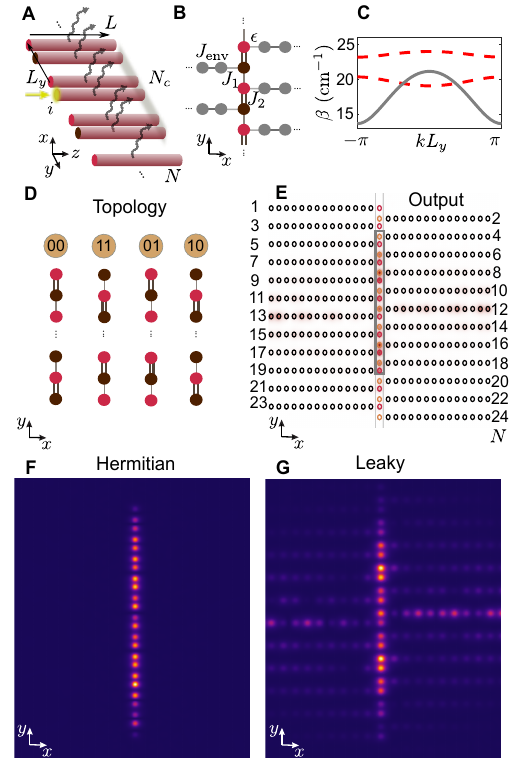}
\caption{ (A) Schematic of a dimerized lattice of single-mode dielectric waveguides with tunable radiative losses and a possible experiment: 
the waveguide indexed by $i$ is excited at the input as indicated by a yellow circle, the intensity distribution is measured in the central area of $N_c$ elements at the output of the sample (the gray rectangle) to generate a dataset for learning the topological properties.
(B) Tight binding model visualization of the photonic lattice in (A).
The red and orange circles depict the main array -- a one-dimensional dimerised SSH-like array of coupled elements. Gray circles illustrate auxiliary arrays constituting leaky channels attached to the main array. The differing dashing between the elements denote different coupling strengths.
(C) Band structures of the main (dashed red lines) and auxiliary (gray solid line) arrays in the designed leaky photonic lattice inscribed in glass. (D) Different configurations of the two edges in a finite lattice.
(E) The output intensity distribution (colored) overlaid with the proposed lattice cross-section. 
(F,G) Intensity distribution, numerically obtained in paraxial modeling at the output facet of the waveguide array for (F) the Hermitian (lossless) lattice and (G) the lattice with leaky channels. 
}
\label{fig:ml1}
\end{figure}

Conventional schemes for predicting topological properties of the modes $\phi_n(\boldsymbol{r}_{\perp})$ based only on measuring intensity profiles require either the large $L$ limit~\cite{PhysRevLett.102.065703,leykam2021probing} or measuring the evolution as a function of $z$~\cite{PhysRevLett.122.193903,PhysRevLett.118.130501}. On the other hand, machine learning approaches can in principle infer topological properties using intensity measurements at a fixed propagation distance~\cite{Zhang_PRL_2018,Rem2019,PhysRevB.102.054107}, at least given access to a sufficient amount of high quality training data.

As a specific example, in the following we consider the leaky Su-Schrieffer-Heeger waveguide lattice shown in Fig.~\ref{fig:ml1}(A), a dimerized array composed of $N$ leaky waveguides with elliptical cross-sections of semi-axes $a_{x,y}$ induced by 
the refractive index perturbations of magnitude $n_A$~\cite{leykam2021probing}. With increasing coupling between the structural elements, 
some supermodes of the lattice become radiative, acquiring a finite lifetime. The radiation losses can be fine-tuned by optimizing the effective potential of the environment and 
radiation channels. This will allow us to study how changes to the input dataset affect the performance of machine learning-based classification of the different topological phases of this lattice. One possible implementation of the radiation channels is by coupling the main array to auxiliary arrays, each consisting of $N_{\text{env}}$ equidistantly spaced single mode waveguides with an index contrast $n_B$, as shown in Fig.~\ref{fig:ml1}(B,D). Examples of feasible parameters close to those employed in the experimental work Ref.~\cite{mukherjee2020observation} are given in Table~\ref{tab:param_full}.

\begin{table}
\centering
\begin{tabular}{l|l} 
Parameter & Value \\ \hline \hline
     $a_y$  &  $5.4 ~\mathrm{\mu m} $\\
     $a_x$  &  $4 ~\mathrm{\mu m}$\\
     $d_1$  &  $17 ~\mathrm{\mu m}$\\
     $d_2$  &  $23 ~\mathrm{\mu m}$\\
     $\rho$  &  $17 ~\mathrm{\mu m}$\\
     $d_\epsilon$  &  $19 ~\mathrm{\mu m}$\\ \hline
    $n_0$  &  $1.47$  \\
    $n_A$  &  $1.2 \times 10^{-3}$  \\
    $n_B$  &  $1.1 \times 10^{-3}$  \\
    $\lambda$ & $1030~\mathrm{nm}$\\
      \hline
\end{tabular}
\caption{Parameters of the designed leaky photonic lattice: 
semiaxes of elliptical single-mode waveguides  $a_{x,y}$; center-to-center distances $d_{1,2}$ between waveguides along the vertical axis; center-to-center distance $\rho$ between waveguides along the horizontal axis. Arrays of auxiliary waveguides are set aside from the main array at a distance $d_\epsilon$. Here, $\lambda$ is the operating wavelength, $n_0$ is the background refractive index of silica glass, 
$n_{A,B}$ are the perturbations of the refractive index inside the waveguides of the main array and arrays of the environment, respectively.}
\label{tab:param_full}
\end{table}

Provided only one band of the main array overlaps with the dispersion curve of side-coupled leaky channels, 
an initially localised excitation with a broad transverse wavenumber spectrum would undergo gradual radiation and decay during propagation.
Therefore, only the top branch will remain populated after a certain propagation distance, making it possible to calculate the topological invariant of the band using the projector of the output field distribution following the method used in Ref.~\cite{leykam2021probing}. 
However, this recipe generally requires knowing the complex-valued field, whereas phase retrieval could be a challenging task. We will demonstrate the possibility to unravel topology of the sample lattice based solely on the output intensity profile in a roughly center-positioned floating window with the use of machine and deep learning methods.

\begin{figure}[t]
	\centering
	\includegraphics[width=8.4cm]{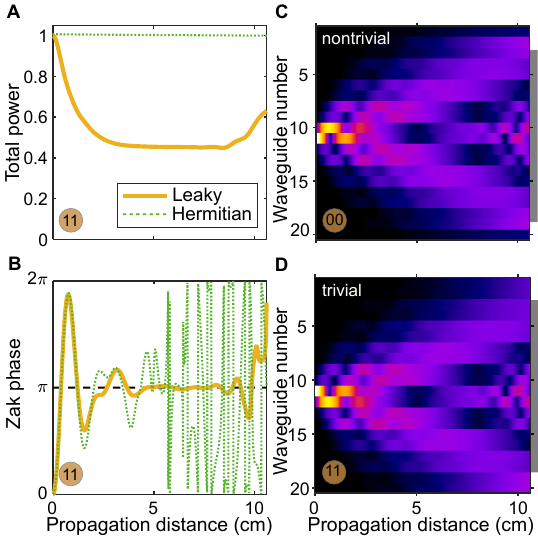}
	\caption{ (A,B) Evolution characteristics of the field in the main array in the lattice with fixed parameters obtained in the TBM 
 of the nontrivial SSH array with (gold curves) and without (green curves) leaky channels. The Zak phase at $z>4$~cm converges to the quantised $\pi$ value, provided $N_{\text{env}}= 14$ elements in leaky channels. 
(C,D) Field evolution in $N$ elements of the main array assembled in a nontrivial (C) and trivial (D) configuration with fixed parameters of the lattice. The gray line on the right side marks the area of $N_c$ central waveguides, the intensity of which is fed to the input of the neural network. }
	\label{fig:ml_2}
\end{figure}

To simplify propagation simulations, 
we constructed the tight binding model (TBM) corresponding to the schematic in Fig.~\ref{fig:ml1}(B) and determined 
parameters of the effective Hamiltonian
in compliance with the paraxial modeling, 
\begin{subequations} \label{eq:tight_binding_eqs_1}
\begin{gather}
i\frac{\partial \psi_m}{\partial z} = \hat{H}_0 \psi_m+ \epsilon c_{m1},\\
i\frac{\partial c_{m1}}{\partial z} = \Delta c_{m1} + \epsilon \psi_m + J_{\text{env}}c_{m2},\\
i\frac{\partial c_{ml}}{\partial z} = \Delta c_{ml} +  J_{\text{env}}(c_{ml-1} + c_{ml + 1}), \quad l=2,...N_{\text{env}}, 
\end{gather}
\end{subequations}
where $\psi_{m}$ and $c_{ml}$ are the amplitudes of the optical field in the main array and in the leaky channels, respectively, $\hat{H}_0$ is the $N \times N$ Hamiltonian of the 
main array, made of the alternating nearest-neighbor (NN) coupling coefficients $J_{1,2}$, $\epsilon$ is the coupling strength between the main array and the environment, $J_{\text{env}}$ is the NN hopping coefficient in leaky channels, and $\Delta$ is a detuning of the propagation constants. 

The dispersion characteristics of the disconnected (at $\eps=0$) uniform lattices representing the main (SSH) array and environment (env) are given by
\begin{subequations}
\begin{align}
&\beta_{\text{SSH}}(k)=\pm \sqrt{J_1^2+J_2^2+2 J_1 J_2 \cos{k L_y}},\\
&\beta_{\text{env}}(k)=\Delta +  2 J_{\text{env}} \cos{k L_y}.  
\end{align}
\end{subequations}
and plotted in Fig.~\ref{fig:ml1}(C). As deliberately ensured by design, the environmental array's 
dispersion curve fully intersects the lower band of the SSH lattice, meaning that only the lower band becomes lossy. Given dimerization, the main array is known to be topologically nontrivial for $J_1<J_2$ and topologically trivial for $J_1>J_2$. 

\begin{table}[t] 
\centering
\begin{tabular}{l|l}  
Parameter & Range \\ \hline \hline
     $J_k$  &  $[1.5;2]$\\
     $J_p$  &  $[0.4;0.6]$\\
     $J_{env}$   &  $[1.7;2]$\\
     $\epsilon$  &  $[0.8;1]$\\
     $\Delta$    &  $[-3.3;-3.5]$\\
     $L$         &  $[2.6;10.6]$\\
     $N$ &  $[20;26]$\\
     $N_{env}$   & $[14]$\\
     $N_c$   & $16$ \\
\end{tabular}
\caption{Ranges of parameters 
used in data set preparation. 
Average values of the listed TBM parameters correspond to the physical quantities in Table~\ref{tab:param_full}, 
as established in paraxial modeling. $k=2,~p=1$ in the nontrivial lattice ($J_1<J_2$), and $k=1,~p=2$ in the trivial lattice ($J_1>J_2$). While preparing the datasets, $J_{1,2}$ were uniformly sampled from within the specified intervals for each vector.
}
\label{tab:param_calcs}
\end{table}

To prepare a dataset, the TBM equations~\eqref{eq:tight_binding_eqs_1} were solved numerically. At the input, we excite a single waveguide designated as $i$ in Fig.~\ref{fig:ml1}(A). The use of a single-element input is justified by its wide spectrum, which allows populating both bands of the lattice.
By iterating over parameters of the photonic lattice in the ranges indicated in Table~\ref{tab:param_calcs}, we accumulated 
data for the analysis of topology of the main array. 
We take into account that the lattice ends can be different, so that $N$ can be odd. We select a sample window composed of a finite number $N_c$ of the central waveguides in the main array. 
Thereby, we aim to solve the classification problem for a finite lattice sample, i.e., to distinguish between different configurations of the two edges based on the intensity distribution measured at the output of $N_c$ central waveguides. The edge of the SSH main array can be either trivial
(0) or non-trivial (1), depending on the lattice termination by strong or weak bond.
The nontrivial edge supports a midgap topological edge state. 
This yields four classes in total: 00, 11, 10, 01.
The four possible configurations are visualized in Fig.~\ref{fig:ml1}(E):  
01 (left trivial, right non-trivial), 11 (left non-trivial, right non-trivial), 10 (left non-trivial, right trivial), 00 (left trivial, right trivial). 
Note that such setup of the problem is different from that in  Ref.~\cite{leykam2021probing}, where both edges of the lattice had the same termination. Also, to calculate the field projector, the field distribution over all elements of the main array was used, that is $N_c=N$ with $N$ even.  

Our previous work~\cite{leykam2021probing} presented a
proposal for calculating the topological invariant (Zak phase) for this lattice (of classes 00 or 11) using the field projector of the output distribution.
This procedure is summarized in Fig.~\ref{fig:ml_2}. By analyzing the complex-valued field distribution [note Fig.~\ref{fig:ml_2}(C,D) only shows the intensity], we compute the Zak phase, which asymptotically approaches $\pi$ in the nontrivial configuration [see Fig.~\ref{fig:ml_2}(A)], provided the leaky channels are introduced. 
At distances $4~\mathrm{cm}<z<9~\mathrm{cm}$ the upper band is completely depopulated as a result of leakage. This depopulation is also evident in the total wavepacket norm, which converges towards $1/2$.
However, when the propagation distance is increased beyond $z>9~\mathrm{cm}$, reflections occur from the ends of the finite environment array and the main lattice, resulting in an increase in the total wavepacket norm [see Fig.~\ref{fig:ml_2}(B)], rendering the 
method inapplicable. Thus, accurate reconstruction of the topological invariant requires either a large lattice or a well-controlled propagation length to avoid reflections off the ends.

\section{Unsupervised learning}
\label{sec:tsne}

 \begin{figure}
	\centering
\includegraphics[width=8.4cm]{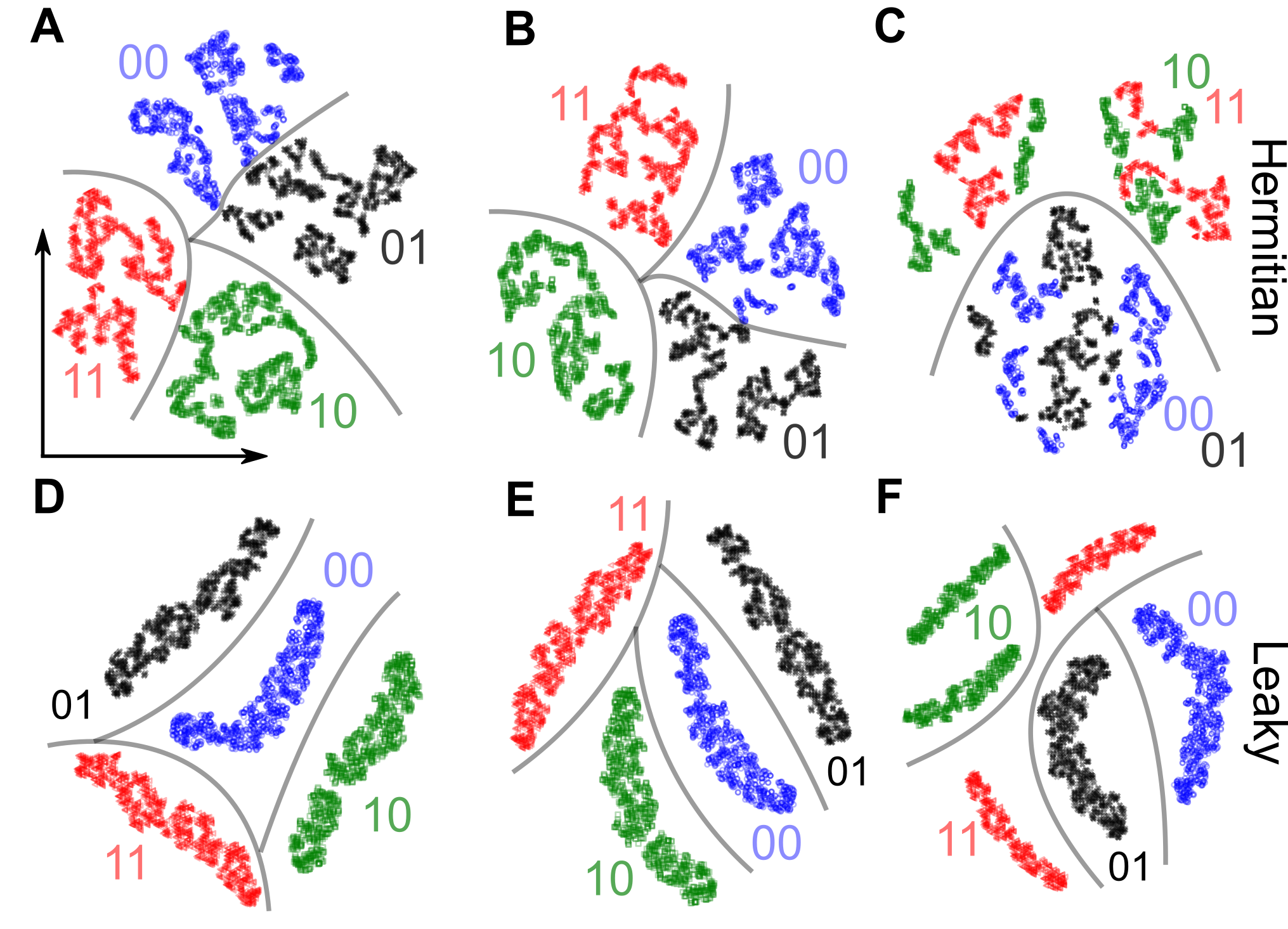}
\caption{ t-SNE maps of the system having 4 topological classes depending on its 2 edges: (A-C) Hermitian lattice, (D-F) lattice with leaky channels. The waveguide excited at the input is indexed by $i$.  
(A,B,D,E) correspond to the case of single-waveguide excitation: (A,D) $i=11$ is odd, 
(B,E) $i=12$ is even, (C,F) the excited waveguide is randomly chosen 
within a dimer. For each point in the two-dimensional parameter space 
there is a corresponding intensity distribution vector of dimension $N=22$ (or $N=23$), depending on the topological class. The four classes are color-coded:  00 (blue), 11 (red), 10 (green), 01 (black). 
}
\label{fig:ml_tsne}
\end{figure}

 \begin{figure}
	\centering
\includegraphics[width=8.4cm]{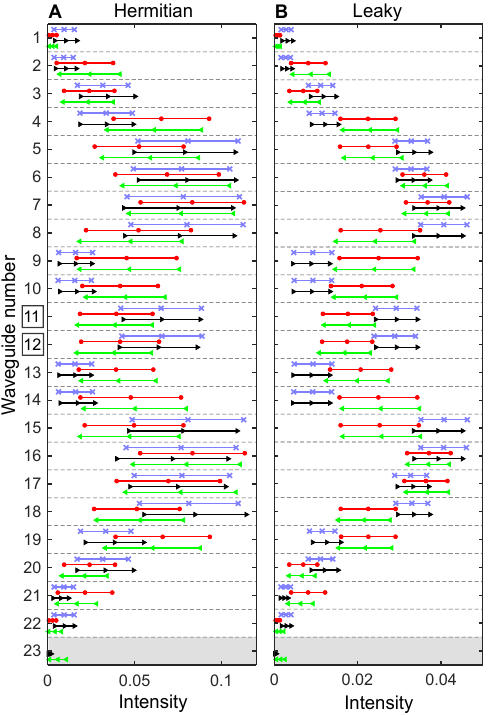} 
\caption{
Statistical characteristics of intensity distributions in waveguides. 
The datasets were prepared for the Hermitian (A) 
and leaky (B) 
cases assuming two possible positions $i=11,12$ of the initial excitation at $L=7.6$~cm.
The mean value is indicated by markers in the middle of horizontal lines, while the standard deviation is represented by the borders of the lines.
The classes are color-coded: 00 (blue squares), 11 (red circles), 01 (black right-facing triangles), 10 (green left-facing triangles). 
The total number of waveguides $N$ is 22 (even) for classes 00 and 11, and 23 (odd) for classes 01 an 10. 
}
\label{fig:ml_stat}
\end{figure}

To begin, we perform the preliminary analysis of the prepared datasets using the t-SNE (t-distributed Stochastic Neighbor Embedding) method. t-SNE is a nonlinear dimensionality reduction algorithm which learns a low-dimensional embedding of the input data; points within the input data set that are close to each other will remain close to each other in the embedded space. Ideally, a vector will be most similar to others obtained from the same lattice configuration, resulting in visible clustering in the low-dimensional embedding.

In this approach, we work with the intensity distribution within $N_c=N$ elements ($N=$ 22 or 23, to be more specific), and assume that the pumped waveguide can be shifted from the center of the lattice. 
Figure~\ref{fig:ml_tsne} shows 
t-SNE maps of the system with fixed $L = 7.6$~cm, $N =22 \: (23)$ and two different positions of the initially excited waveguide. 
In the Hermitian case (leakage disabled), the different classes become mixed up in the embedded space; whereas in the case of a
lattice with leaky channels, they do not. This qualitatively agrees with the theory in  Ref.~\cite{leykam2021probing}, specifically that the different phases will exhibit distinct intensity distributions in their bulk. 

However, as soon as we introduce uncertainty, such as the position of the initial excitation, the topological classes are no longer clearly separable: in the Hermitian case different classes become mixed up [Fig.~\ref{fig:ml_tsne}(C)], whereas in the leaky lattice too many clusters are obtained [Fig.~\ref{fig:ml_tsne}(F)]. Consequently, unsupervised methods are no longer applicable. 

Figure~\ref{fig:ml_stat} presents the statistic analysis of the data used for (C,F) panels of Fig.~\ref{fig:ml_tsne}. 
This visualization shows that classes 01 and 00, 10 and 11 can be grouped pairwise.  
However, the classes with dissimilar edge topologies (01 and 10) are differentiated from the classes with the identical edge topologies (00 and 11) by odd $N$, 
due to distinct input vector lengths (the 23th waveguide 
for which case is shown shaded). 
This postprocessing 
also reveals significant overlaps of the intensity bars for 00 and 11 classes in each waveguide of the Hermitian SSH lattice, while the bars overlap less in the leaky lattice forming shifted dimerized patterns, 
a feature to be noticed by the neural network. 

\section{Supervised learning}
\label{sec:supervised}

For supervised classification {of the four topological classes}, 
we apply machine (K-Nearest Neighbors (KNN), Support Vector Machine (SVM), Decision Tree) 
and deep (Multi-layer Perceptron (MLP), Convolutional Neural Network (CNN)) learning methods (see details in the Supplementary Materials, Section~III). 
%
The numerical experiments were carried out with varying parameters: propagation distance $L$, 
total number of waveguides $N$, 
number of the central waveguides in a sample window $N_c$. 
The input waveguide $i$ can be shifted by 1 from the center of the array, according to the expression $\mathrm{ceil} (N/ 2+l)$, where $l$ can be 0 or 1. 
For each $L$ we obtain a dataset of 32,000 intensity vectors.
Accordingly by a parameter, subsets from the whole data set can be grouped. Let us examine the accuracy of classification depending on different parameters. The metric we use for this non-binary classification problem is the accuracy, defined as the percentage of correct model predictions, 
\begin{equation}
\begin{gathered}
\text{Accuracy} = \frac{\sum_{i=1}^n \mathbb{1}[p_i=y_i]}{n},
\label{eq:acc_formula}
\end{gathered}
\end{equation}
where $p_i$ and $y_i$ are the predicted and the correct answer, respectively, and $\mathbb{1}$ is an indicator function equal to one if the condition is met and zero otherwise. 

\begin{figure}
	\centering
	\includegraphics[scale=1]{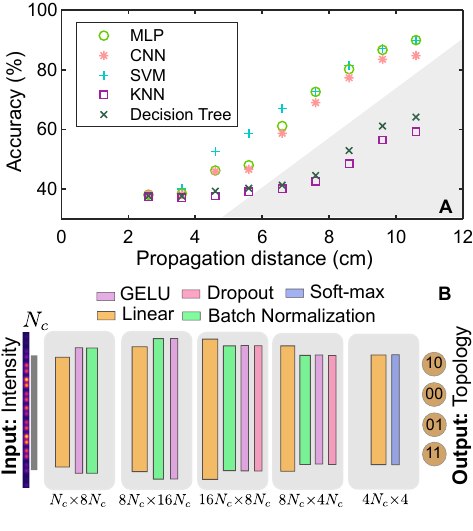}
	\caption{ (A) Accuracy of supervised learning methods as a function of the propagation distance $L$. (B) Scheme of the convolutional neural network, which takes the intensity distribution at $z=L$ as the input and determines topology of the lattice edges, $N_c=16$. 
  }
	\label{fig:ml_4}
\end{figure}

Figure~\ref{fig:ml_4}(A) illustrates how the accuracy of the supervised learning techniques varies with the parameter $L$. The accuracy  increases as the propagation distance increases. 
When the value of $L$ is small, theoretical predictions cannot distinguish between different topological phases, and all methods show similar accuracy plateaus in their graphs.
Further, the accuracy of machine learning methods increases with increasing $L$, see Fig.~\ref{fig:ml_4}(A).
At the same time, the theoretical curve for the Zak phase in the nontrivial case ceases to converge to the quantized invariant value $\pi$ for $L=10.6$~cm [see Fig.~\ref{fig:ml_2}(A)], while the power in the main array tends to grow and exceeds one half [see Fig.~\ref{fig:ml_2}(B)]. This is explained by reflection from the boundaries of leaky channels, as the field returns back to the main array. The requirement to know both the intensity and phase at the output in the method of Ref.~\cite{leykam2021probing} is replaced by statistical information from dynamics, but only intensity distributions at fixed $L$.

Machine learning methods perform better for larger $L$. This may be due to the fact that, as soon as the radiation reaches the edges, to distinguish the trivial case from the non-trivial one, we can consider not only bulk properties but also the edges themselves, and machine learning methods allow us to take this effect into account.
 For instance, the trivial and non-trivial cases are even visually distinguishable in the dynamics shown in Fig.~\ref{fig:ml_2}(C,D): in the non-trivial case the bulk modes poorly couple to 
 waveguides at the edges. Note that if we increase the number of auxiliary waveguides $N_{\text{env}}$, the theoretical power curve will exhibit convergence to $0.5$, but the reflection off the main array edges will still manifest at larger propagation distances. 
Thus, neural network methods are applicable in a wider range  of cases than the theoretical scheme based on the projector calculation. 

\begin{figure} [b]
	\includegraphics[width=8.4cm]{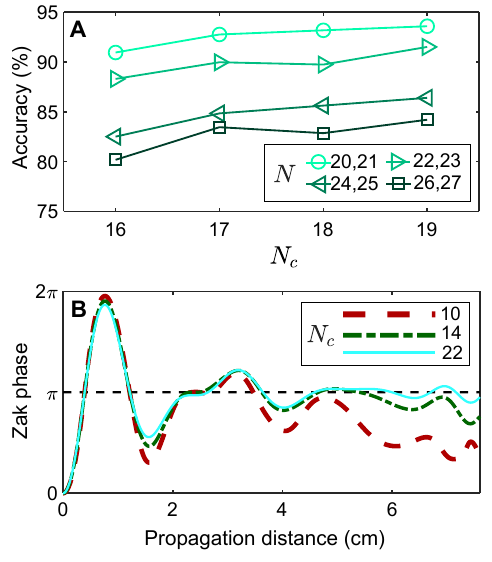}
	\caption{(A) Accuracy of classification by deep learning methods depending on parameters: the total number of waveguides $N$ and the number of the central waveguides $N_c$ involved in the training. (B) Theoretical dependence of the Zak phase on the propagation distance and $N_c$ in the nontrivial lattice of $N=22$ elements.
}
	\label{fig:ml_AcNL} 
\end{figure}

Based on results summarised in Fig.~\ref{fig:ml_4}(A), we conclude that classical machine learning methods show lower accuracy compared to neural networks and support vector machine (SVM).
One of the two most promising models, the MLP method, was chosen for 
more thorough examination in Fig.~\ref{fig:ml_AcNL}.

As noted above, training was held using $N_c<N$ central waveguides. Figure~\ref{fig:ml_AcNL}(A) shows the dependence of the classification accuracy on the number of central waveguides while in training batches all $L$ were involved. In the initially proposed theoretical scheme, we calculated the field projector for $N_c=N$ elements, but we can formally calculate it for any $N_c<N$, as shown in Fig.~\ref{fig:ml_AcNL}(B). The Zak phase is seen to converge better to the correct quantised value for larger $N_c$, and this condition is also necessary to increase the accuracy of machine learning algorithms: in Fig.~\ref{fig:ml_AcNL}(B) the precision increases as the $N_c/N$ ratio increases. 

\begin{figure}
	\includegraphics[width=8.4cm]{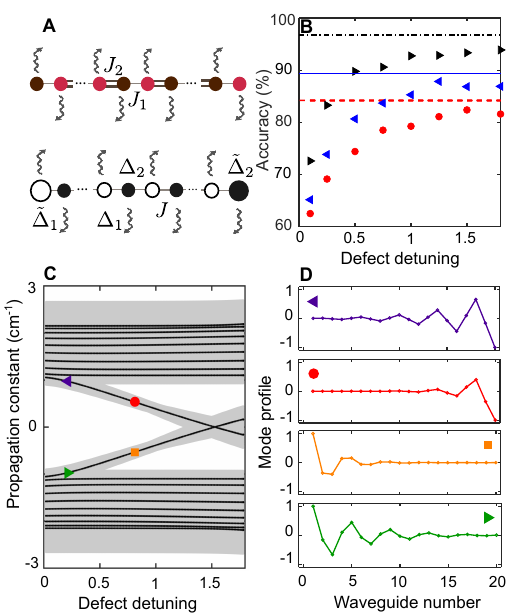} 
	\caption{ 
(A) Schematics of the topological (upper row) dimerized array and the non-topological (lower row) dimer lattice with defect potentials $\tilde{\Delta}_{1,2}$ at the edges.
(B) The accuracy of the neural network trained for the non-topological case for different values of the edge defect detuning $q_1$, introduced as $\tilde{\Delta}_{1,2}=\Delta_{1,2}(1-q_1)$, and different propagation distances $L=7.6$~cm (red dots), $L=8.6$~cm (blue left-facing triangles), $L=10.6$~cm (black right-facing triangles). For comparison, the colored horizontal lines depict the accuracy in the topological case for the corresponding $L$. (C) The 
band structure of the finite non-topological lattice depending on the defect detuning, at the fixed number of elements within the main array $N=22$. The shading 
shows bands for all possible coupling coefficients, $J$, and detunings, $\Delta_{1} = - \Delta_{2}$, that were utilized to generate the datasets. 
(D) Profiles of the modes bound to the ends of the non-topological lattice. Colors and shapes of the markers in (C) in the representative spectral positions correspond to the %
profiles in (D).
}
\label{fig:ml_vs} 
\end{figure}

To better understand the performance of the supervised classification approach at distinguishing the different edge types, we compare topological SSH lattice with even number of elements and its non-topological counterpart, where dimerization is stipulated by the alternating 
difference in propagation constants 
($\Delta_1$ and $\Delta_2 = - \Delta_1$), whereas the coupling between neighboring elements is uniform and equal to $J$, as schematically shown in Fig.~\ref{fig:ml_vs}(A). To prepare the corresponding datasets, parameters of the non-topological lattice ($\Delta_1$ and $J$) are chosen 
such that its band structure coincides with the topological one (see Supplementary Materials, Section I). We introduce  trivial edge defects as detunings of the propagation constant in the edge elements. Thereby, the defect potential for the left end is $\tilde{\Delta}_1=\Delta_1(1-q_1)$, whereas the defect potential for the right end is $\tilde{\Delta}_2=\Delta_2(1-q_2)$. We compare the accuracy of the neural network at three propagation distances [see Fig.~\ref{fig:ml_vs}(B)] for the topological SSH array and non-topological array with the edge defects 
in distinguishing the two classes: both edges either support  confined solutions (class 11) or not (class 00).
We find that for small amplitudes of the defect 
the accuracy for the case of the non-topological lattice is small compared to the topological one, since the defect is not connected to its bulk properties (unlike in the topological case), but bulk modes also change when the defect amplitude becomes large, leading to an increase in the model accuracy.

\section{Disorder and transfer learning}
\label{sec:disorder}

Transfer learning refers to the use of a model trained on one set of data to make accurate predictions on a new task. Here we consider the performance of models trained on ideal data in classifying data obtained from different model parameters.  If the quality metric falls slightly, we can conclude that the model has a generalization ability. This is particularly important in the context of nanophotonic circuits, where inevitable disorder will lead to sample-to-sample variations of device parameters. 

First, we note that the generalization ability is not observed for the parameter $L$, and the accuracy drops significantly when testing on $L$ different from the propagation distance used for the training data. On the other hand, we observe generalization over some $N$, that corresponds to attaching dimers to both edges of the main array, stipulated by the fact that such an addition of elements does not qualitatively change the topology of the lattice (see the cross-validation control map for parameter $N$ in Supplementary Materials, Section IV).

Next, we examine a transfer learning approach that allows for the reuse of pretrained models at a fixed propagation distance of $L = 10.6$~cm [referring to the last point in Fig.~4(A)] on models with disorder. We introduce perturbations into the SSH Hamiltonian coefficients of two types: off-diagonal disorder in the inter-site coupling strengths and on-site disorder in the propagation constants. Incorporating disorder involves adding random variables to the coefficients of the Hamiltonian.
For example, the off-diagonal disorder perturbs each coupling coefficient by the random variable $ l \langle d \rangle  \mathrm{mean} (J_1,J_2)$, where $l$ is uniformly distributed in the range $[-1/2, 1/2]$ and $\langle d \rangle$ is the disorder strength. This is a chiral type disorder in the sense that the Hamiltonian describing the disordered system respects the chiral symmetry, thus its topological edge states will remain at zero energy.
We train the neural
network using a non-disordered array and test it on the disordered lattice. We have identified a range of disorder strengths in which the previously trained neural network can operate with high confidence.

\begin{figure}[tb!]
	\includegraphics[width=8.4cm]{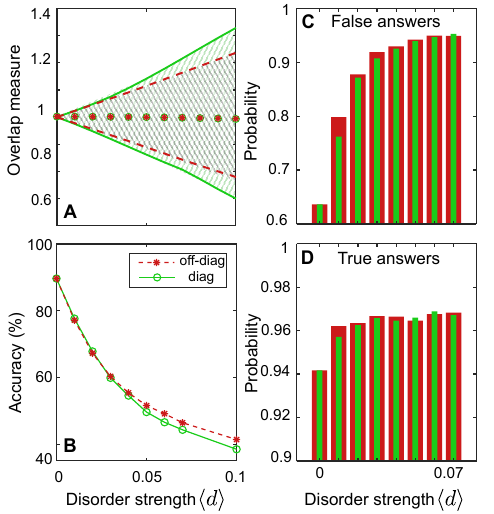}
	\caption{
 (A) Overlap measure variation induced by the disorder: shaded areas are ranges of variance due to disorder over an ensemble of 4000 disorder realizations (green is for diagonal disorder, gray for off-diagonal disorder), asterisks and dots are mean values. All parameters of the lattice are fixed.
 (B) Transfer learning for the disordered lattice. We train neural network in the absence of disorder $\langle d \rangle=0$ and test the prediction accuracy for different values of disorder. All parameters of the lattice are varied according to Table~\ref{tab:param_calcs}. (C,D) Probability assigned to false (C) and true (D) answers of the neural network for different values of disorder (green bars are for diagonal disorder, gray bars are for off-diagonal disorder).}
	\label{fig:ml_transfer} 
\end{figure}

To quantify the impact of the disorder on the data, we compute the similarity between the output intensities. Specifically, we compute the output fields $\psi_m (\langle d \rangle,i)^{1,2}$, where the indices 1 and 2 correspond to diagonal and off-diagonal disorders, respectively, and $i$ represents the number of the specific disorder realization.
We then introduce the intensity 
overlap as 
$\mathcal{O}^{1,2} (\langle d \rangle,i) = \sum_m |\psi_m(\langle d \rangle,i)^{1,2})|^2 \cdot |\psi_m^0|^2$, where summation is taken over waveguides of the main array and $\psi_m^0$ is the output distribution in the disorder-free case. This overlap measures the similarity between the two distributions. It is a useful quantity to study the effect of disorder on the output of a system, as it allows us to quantify how much the output changes due to disorder.
To plot the overlap measure, we calculate  $\mathcal{O}^{1,2} (\langle d \rangle,i)$ over 4000 disorder realizations for each of the values of $\langle d \rangle$.
To standardize the plotted functions, we divide them by the value of $\mathcal{O}^{1,2} (\langle d \rangle,i)$ when $\langle d \rangle$ is zero. This normalization process allows us to compare the variability of the overlap measure across different scenarios.
The dotted areas in Fig.~\ref{fig:ml_transfer}(A) represent the corresponding ranges. Note, we have rescaled the diagonal disorder strength $\langle d_{\mathrm{diag}} \rangle = 4 \langle d_{\mathrm{off-diag}} \rangle$ such that for a given $\langle d \rangle$ the two forms of disorder have a similar effect on the overlap measure.

To demonstrate transfer learning for disordered arrays, we train the neural network using a non-disordered array and test it for the disordered lattice [see Fig.~\ref{fig:ml_transfer}(B)], the ranges of parameters as in Table~\ref{tab:param_calcs}. The accuracy curves are similar for both types of disorder, showing a decrease in accuracy as the disorder amplitue increases. Expanding the range of the overlap measure results in a significant change in the output intensity, which ultimately leads to a sharp decline in the classification accuracy.

To estimate confidence of the trained neural network, we 
study the output of the last layer [see Fig.~\ref{fig:ml_2}(C)] in detail. 
Softmax function returns probabilities of four classes. Here we fix the class 00 
(both ends are trivial), but the results are comparable for the other classes as well. If the model assigns a high probability to a particular class, it is more confident in that prediction than if it assigns a lower probability.

We create a set of test vectors for each disorder amplitude and select vectors that have the highest probability of belonging to class 00.
If this vector indeed belongs to class 00, we label the probability as true; otherwise, it is labeled as false.
And then we average false and true answers to plot Fig.~\ref{fig:ml_transfer}(C,D). Interestingly, as the accuracy of the neural network decreases, its level of certainty in both accurate and inaccurate responses increases. In other words, the neural network will more confidently give the wrong answer as the disorder strength is increased, indicating that the fabrication disorder can act as an adversarial perturbation.

\section{Conclusion}
\label{sec:conclusion}

We have studied the performance of a variety of machine learning techniques at distinguishing different topological phases of leaky photonic lattices using measurements of the bulk intensity profile after a fixed propagation distance. First, we found that uncertainty in the initial conditions (such as the excited waveguide) reduces the quality of unsupervised clustering, leading to either mixing between different classes or the prediction of too many classes. We then compared the performance of a different supervised learning methods, finding that high accuracy can be achieved for sufficiently large propagation distances. The classification accuracy can be further improved by increasing the number of bulk waveguide intensities used. Finally, we studied the transfer learning ability of neural network-based classifiers. While the accuracy drops significantly if the network is trained on data obtained using a different propagation distance, the networks can accurately classify data from systems with sufficiently weak disorder, thus avoiding extensive training on each new system. Our approach for classifying lattices based on incomplete measurements can be further developed to solve a more general problem of reconstruction of the lattice Hamiltonian with some a priori knowledge of its symmetries in various fields including photonics, condensed matter physics, and quantum computing.

\section*{Acknowledgements}
The authors acklnowledge useful discussions with Clemens Gneiting, Alexey Horkin and Nikita Kulikov. E.S. and L.S. are supported in part by the MSHE under project No. 0729-2021-013. E.S. thanks the Foundation for the Advancement of Theoretical Physics and Mathematics "BASIS" (Grant No. 22-1-5-80-1). D.L. acknowledges support from the National Research Foundation, Singapore and A*STAR under its CQT Bridging Grant. F.N. is supported in part by: Nippon Telegraph and Telephone Corporation (NTT) Research, the Japan Science and Technology Agency (JST) [via the Quantum Leap Flagship Program (Q-LEAP), and the Moonshot R\&D Grant Number JPMJMS2061], the Asian Office of Aerospace Research and Development (AOARD) (via Grant No. FA2386-20-1-4069), and the Foundational Questions Institute Fund (FQXi) via Grant No. FQXi-IAF19-06. D.S. acknowledges support from the Australian Research Council (FT230100058) and the Japan Society for the Promotion of Science under the Postdoctoral Fellowship Program for Foreign Researchers. 


\begin{thebibliography}{100}
\newcommand{\enquote}[1]{``#1''}

\bibitem{RevModPhys.91.045002}
G.~Carleo, I.~Cirac, K.~Cranmer, L.~Daudet, M.~Schuld, N.~Tishby,
  L.~Vogt-Maranto, and L.~Zdeborov\'a, \enquote{Machine learning and the
  physical sciences,} Rev. Mod. Phys. \textbf{91}, 045002 (2019).

\bibitem{Wiecha2020}
P.~R. Wiecha and O.~L. Muskens, \enquote{Deep learning meets nanophotonics: A
  generalized accurate predictor for near fields and far fields of arbitrary 3d
  nanostructures,} Nano Letters \textbf{20}, 329--338 (2020).

\bibitem{Chen2022}
M.~Chen, R.~Lupoiu, C.~Mao, D.-H. Huang, J.~Jiang, P.~Lalanne, and J.~A. Fan,
  \enquote{High speed simulation and freeform optimization of nanophotonic
  devices with physics-augmented deep learning,} ACS Photonics \textbf{9},
  3110--3123 (2022).

\bibitem{Melati2019}
D.~Melati, Y.~Grinberg, M.~Kamandar~Dezfouli, S.~Janz, P.~Cheben, J.~H. Schmid,
  A.~S{\'a}nchez-Postigo, and D.-X. Xu, \enquote{Mapping the global design
  space of nanophotonic components using machine learning pattern recognition,}
  Nature Communications \textbf{10}, 4775 (2019).

\bibitem{malkiel_mrejen_wolf_suchowski_2020}
I.~Malkiel, M.~Mrejen, L.~Wolf, and H.~Suchowski, \enquote{Machine learning for
  nanophotonics,} MRS Bulletin \textbf{45}, 221--229 (2020).

\bibitem{SoBadloeNohBravoAbadRho+2020+1041+1057}
S.~So, T.~Badloe, J.~Noh, J.~Bravo-Abad, and J.~Rho, \enquote{Deep learning
  enabled inverse design in nanophotonics,} Nanophotonics \textbf{9},
  1041--1057 (2020).

\bibitem{Wiecha:21}
P.~R. Wiecha, A.~Arbouet, C.~Girard, and O.~L. Muskens, \enquote{Deep learning
  in nano-photonics: inverse design and beyond,} Photon. Res. \textbf{9},
  B182--B200 (2021).

\bibitem{LiuLiuZhouLiYouQiuWu+2023+1943+1955}
G.-X. Liu, J.-F. Liu, W.-J. Zhou, L.-Y. Li, C.-L. You, C.-W. Qiu, and L.~Wu,
  \enquote{Inverse design in quantum nanophotonics: combining
  local-density-of-states and deep learning,} Nanophotonics \textbf{12},
  1943--1955 (2023).

\bibitem{doi:10.1080/23746149.2022.2046156}
J.~Yun, S.~Kim, S.~So, M.~Kim, and J.~Rho, \enquote{Deep learning for
  topological photonics,} Advances in Physics: X \textbf{7}, 2046156 (2022).

\bibitem{Pilozzi2018}
L.~Pilozzi, F.~A. Farrelly, G.~Marcucci, and C.~Conti, \enquote{Machine
  learning inverse problem for topological photonics,} Communications Physics
  \textbf{1}, 57 (2018).

\bibitem{ozawa2019topological}
T.~Ozawa, H.~M. Price, A.~Amo, N.~Goldman, M.~Hafezi, L.~Lu, M.~C. Rechtsman,
  D.~Schuster, J.~Simon, O.~Zilberberg \emph{et~al.}, \enquote{Topological
  photonics,} Reviews of Modern Physics \textbf{91}, 015006 (2019).

\bibitem{Rodriguez-Nieva2019}
J.~F. Rodriguez-Nieva and M.~S. Scheurer, \enquote{Identifying topological
  order through unsupervised machine learning,} Nature Physics \textbf{15},
  790--795 (2019).

\bibitem{PhysRevLett.124.226401}
M.~S. Scheurer and R.-J. Slager, \enquote{Unsupervised machine learning and
  band topology,} Phys. Rev. Lett. \textbf{124}, 226401 (2020).

\bibitem{PhysRevLett.125.127401}
E.~Lustig, O.~Yair, R.~Talmon, and M.~Segev, \enquote{Identifying topological
  phase transitions in experiments using manifold learning,} Phys. Rev. Lett.
  \textbf{125}, 127401 (2020).

\bibitem{PhysRevB.102.134213}
Y.~Che, C.~Gneiting, T.~Liu, and F.~Nori, \enquote{Topological quantum phase
  transitions retrieved through unsupervised machine learning,} Phys. Rev. B
  \textbf{102}, 134213 (2020).

\bibitem{PhysRevLett.130.036601}
Y.~Long and B.~Zhang, \enquote{Unsupervised data-driven classification of
  topological gapped systems with symmetries,} Phys. Rev. Lett. \textbf{130},
  036601 (2023).

\bibitem{PhysRevResearch.2.023283}
Y.~Zhang, P.~Ginsparg, and E.-A. Kim, \enquote{Interpreting machine learning of
  topological quantum phase transitions,} Phys. Rev. Res. \textbf{2}, 023283
  (2020).

\bibitem{Zhang2022}
H.~Zhang, S.~Jiang, X.~Wang, W.~Zhang, X.~Huang, X.~Ouyang, Y.~Yu, Y.~Liu,
  D.-L. Deng, and L.-M. Duan, \enquote{Experimental demonstration of
  adversarial examples in learning topological phases,} Nature Communications
  \textbf{13}, 4993 (2022).

\bibitem{8848487}
A.~Blanco-Redondo, \enquote{Topological nanophotonics: {T}oward robust quantum circuits,} Proceedings of the IEEE \textbf{108}, 837--849 (2020).

\bibitem{https://doi.org/10.1002/adpr.202100010}
D.~T.~H. Tan, \enquote{Topological silicon photonics,} Advanced Photonics
  Research \textbf{2}, 2100010 (2021).

\bibitem{PhysRevResearch.4.033222}
J.~Gao, Z.-S. Xu, D.~A. Smirnova, D.~Leykam, S.~Gyger, W.-H. Zhou,
  S.~Steinhauer, V.~Zwiller, and A.~W. Elshaari, \enquote{Observation of
  {A}nderson phase in a topological photonic circuit,} Phys. Rev. Res.
  \textbf{4}, 033222 (2022).

\bibitem{PhysRevLett.102.065703}
M.~S. Rudner and L.~S. Levitov, \enquote{Topological transition in a
  non-{H}ermitian quantum walk,} Phys. Rev. Lett. \textbf{102}, 065703 (2009).

\bibitem{leykam2021probing}
D.~Leykam and D.~A. Smirnova, \enquote{Probing bulk topological invariants
  using leaky photonic lattices,} Nature Physics \textbf{17}, 632--638 (2021).

\bibitem{PhysRevLett.122.193903}
Y.~Wang, Y.-H. Lu, F.~Mei, J.~Gao, Z.-M. Li, H.~Tang, S.-L. Zhu, S.~Jia, and
  X.-M. Jin, \enquote{Direct observation of topology from single-photon
  dynamics,} Phys. Rev. Lett. \textbf{122}, 193903 (2019).

\bibitem{PhysRevLett.118.130501}
V.~V. Ramasesh, E.~Flurin, M.~Rudner, I.~Siddiqi, and N.~Y. Yao,
  \enquote{Direct probe of topological invariants using {B}loch oscillating
  quantum walks,} Phys. Rev. Lett. \textbf{118}, 130501 (2017).

\bibitem{Zhang_PRL_2018}
P.~Zhang, H.~Shen, and H.~Zhai, \enquote{Machine learning topological
  invariants with neural networks,} Phys. Rev. Lett. \textbf{120}, 066401
  (2018).

\bibitem{Rem2019}
B.~S. Rem, N.~K{\"a}ming, M.~Tarnowski, L.~Asteria, N.~Fl{\"a}schner,
  C.~Becker, K.~Sengstock, and C.~Weitenberg, \enquote{Identifying quantum
  phase transitions using artificial neural networks on experimental data,}
  Nature Physics \textbf{15}, 917--920 (2019).

\bibitem{PhysRevB.102.054107}
N.~L. Holanda and M.~A.~R. Griffith, \enquote{Machine learning topological
  phases in real space,} Phys. Rev. B \textbf{102}, 054107 (2020).

\bibitem{mukherjee2020observation}
S.~Mukherjee and M.~C. Rechtsman, \enquote{Observation of {F}loquet solitons in a topological bandgap,} Science \textbf{368}, 856--859 (2020).

\end{thebibliography}

\end{document}